\documentclass{ws-mpla}
\usepackage[super]{cite}
\usepackage{graphicx}
\usepackage{CJK}
\usepackage{indentfirst}
\usepackage{bm}
\usepackage{soul}
\usepackage{xcolor}
\begin{document}

\markboth{Ostergaard, Costa, Sang}{Cosmological Forecasts from BAO in 21cm Intensity Mapping}

\catchline{}{}{}{}{}

\title{Cosmological Forecasts from the Baryon Acoustic Oscillations in 21cm Intensity Mapping}

\author{Benjamin Ostergaard\footnote{
Present address}}

\address{Center for Gravitation and Cosmology, College of Physical Science and Technology, Yangzhou University, Yangzhou, 225009, China\\
benjaminostergaard@proton.me}

\author{Andr\'{e} A. Costa}

\address{Instituto Federal de Educa\c{c}\~{a}o, Ci\^{e}ncia e Tecnologia de Minas Gerais, Campus Itabirito\\
Itabirito, 35450-000, Brazil\\}
\address{Center for Gravitation and Cosmology, College of Physical Science and Technology, Yangzhou University, Yangzhou, 225009, China\\}
\address{College of Physics, Nanjing University of Aeronautics and Astronautics\\
Nanjing 211106, China}
\address{Departamento de F\'{i}sica, Universidade Federal da Para\'{i}ba\\
Caixa Postal 5008, Jo\~{a}o Pessoa 58051-900, Para\'{i}ba, Brazil}

\author{Yu Sang}

\address{Center for Gravitation and Cosmology, College of Physical Science and Technology, Yangzhou University\\
Yangzhou, 225009, China
}

\maketitle

\pub{Received (Day Month Year)}{Revised (Day Month Year)}

\begin{abstract}
In this work we use a template method to extract the scale associated with the Baryon Acoustic Oscillation (BAO) signal in 21cm neutral hydrogen intensity maps. We then forecast the constraints on the standard deviations of cosmological parameters using a Fisher matrix analysis. In order to test this method, we choose the survey configuration for the BINGO telescope. We then estimate the constraints on the BAO shift parameter $\alpha$, which we extract from the 21cm angular power spectrum (APS). In addition, we translate those results into constraints on the final cosmological parameters. As BAO data alone can only constrain the product of the Hubble constant and the sound horizon $H_0r_s$, degeneracies between the variables mean that we can't get useful constraints with BAO data alone. We break these degeneracies by combining the 21cm intensity mapping BAO results with the Cosmic Microwave Background (CMB) covariances obtained by the Planck satellite. In particular, we find that the best forecasts we can get with this combination are on the standard deviations of the Hubble parameter $\sigma_h$, and the dark energy parameters $\sigma_{w_0}$ and $\sigma_{w_a}$. We find $\sigma_h = 0.0055\;(0.8\%)$ in the $\Lambda$CDM model. For the $w$CDM model, we find $\sigma_h = 0.020\;(2.9\%)$ and $\sigma_{w_0} = 0.075\;(7.5\%)$. In the CPL parameterization, we find $\sigma_h = 0.029\;(4.4\%)$, $\sigma_{w_0} = 0.40\;(40\%)$, and $\sigma_{w_a} = 1.7$. Finally, we observe that using the full APS provides stronger constraints than the BAO only, however, it is more susceptible to systematic effects.

\keywords{cosmology; baryon acoustic oscillations; 21cm intensity mapping.}
\end{abstract}

\ccode{PACS Nos.: 98.80.-k}

\section{Introduction}
Baryon acoustic oscillations are sinusoidal oscillations found in the power spectra of both the cosmic microwave background and large-scale structure. They are the result of overdensities in the primordial plasma. In the adiabatic model, the overdensity includes all particle species \cite{pe00300h, Eisenstein_2007}. Photons overpressure the region and drive a spherical sound wave carrying baryons along with it. The cold dark matter does not interact with any other species and thus remains confined in a small radius. After decoupling, the sound wave freezes out as the sound speed decreases rapidly. We are left with a spherical baryon overdensity with a radius $r_s$ and a CDM overdensity at the center of the sphere. These overdensities at both the sphere and its center grow by gravitational instability and provide the seeds for galaxy generation \cite{pe00300h, Eisenstein_2007}.

Our universe is then composed of many superpositions of these spheres. Fortunately, the perturbations behave linearly at high redshifts so we can simply add them all together \cite{Eisenstein_2007}. Of course, the sphere does not have zero thickness. There are three factors that determine its width: Photons leak out of the sound wave producing a phenomenon called Silk damping \cite{1968ApJ...151..459S}; Over time the inertia of the baryons relative to the photons increases thus changing the sound speed and broadening the wave adiabatically \cite{Eisenstein_2007}; Finally, the initial perturbations can interfere with each other. Since galaxies are more likely to form on these spheres, we see a peak in the galaxy 2-point correlation function at about $\sim 150$Mpc in comoving coordinates. This peak in the correlation function corresponds to an oscillation in the power spectrum, the Fourier transform of the correlation function. 

In this paper, we use intensity maps of 21cm line of neutral hydrogen as a tracer for the matter angular power spectrum. The hydrogen atom is one of the simplest objects in atomic physics, simply consisting of a proton and an electron. In its ground 1S state, there are actually two ``hyperfine" states related to the interaction of the proton and electron's magnetic moments. A hydrogen atom in the excited hyperfine state, where both the spins of the proton and electron are aligned, can spontaneously decay into the lower hyperfine state where the electron's spin ``flips" and the two spins become antiparallel. The energy difference between these two states is $\Delta E = 5.9 \times 10^{-6}$eV. Energy conservation leads to the difference being accounted for by a release of a photon with this energy. The corresponding wavelength and frequency of this photon are 21cm and 1420 MHz, respectively. Studies of hydrogen masers have provided very precise measurements of these quantities \cite{PhysRevLett.5.361}. Thus, the 21cm line makes a great, well-understood astronomical probe.

The main difference between intensity mapping and galaxy surveys, the other major low-redshift observation technique, is that while galaxy surveys must resolve each individual galaxy and their redshifts, intensity mapping simply counts the number of photons that land on each pixel of the observation volume. Essentially, intensity mapping surveys are simply digital cameras. The instruments, data, and computing resources required for galaxy surveys such as the Sloan Digital Sky Survey \cite{2000AJ....120.1579Y} are astronomical. Conversely, the instruments required for intensity maps are comparatively cheaper as the telescope light collection area can be smaller. Data analysis is also easier as galaxy distributions can be recovered from an intensity map by calculating and applying the bias to the power spectrum, a much easier task than individually resolving and calculating the redshift of every galaxy in the survey volume.

Because of these many advantages, there are several experiments under way to observe the 21cm signal specifically for the purpose of measuring BAOs. We mention a few: the Canadian Hydrogen Intensity Mapping Experiment (CHIME)\cite{Amiri_2018}, MeerKAT\cite{MeerKLASS:2017vgf}, the Square Kilometer Array (SKA)\cite{2020PASA...37....2W}, and Baryon acoustic oscillations from Integrated Neutral Gas Observations (BINGO)\cite{Battye_2013, 10.1093/mnras/stv2153, BINGO_I}. 

We use specifications for the BINGO project in our analysis. The BINGO telescope is a single fixed-dish radio telescope currently under construction in Para\'{i}ba, Brazil. BINGO is specifically designed to measure the 21cm emission line of neutral hydrogen at large angular scales from redshift $z = 0.13$ to $z = 0.45$. A set of papers have been published describing the BINGO project in detail. BINGO I \cite{BINGO_I} provides an overview of the project. BINGO II \cite{BINGO_II} provides a detailed instrument description. BINGO III \cite{BINGO_III} discusses the optical design and optimization of the focal plane. BINGO IV \cite{BINGO_IV} simulates the mission and the preliminary steps for component separation. BINGO V \cite{BINGO_V} describes the full foreground component separation and bispectrum analysis of the data. BINGO VI \cite{BINGO_VI} provides a mock to test our data analysis on. BINGO VII \cite{BINGO_VII} makes cosmological forecasts using the Fisher matrix method for the total angular power spectrum and assesses the BINGO setup and how it compares to other telescopes. BINGO VIII \cite{BINGO_VIII} discusses the recoverability of the BAO signal from HI intensity maps using the Markov Chain Monte Carlo method.

This paper helps connect the 21cm Fisher forecast given in BINGO VII with the BAO signal extraction discussed in BINGO VIII. Specifically, we show how to extract and use the BAO signal from the 21cm APS and use it to make forecasts on cosmological parameters using the specifications for the BINGO telescope thus going beyond both papers. In Section 2, we derive the 21cm angular power spectrum we use in our analysis. In particular, we explain how we rewrite the APS given in BINGO VII in terms of the 3D power spectrum $P(k)$ and extract the BAO shift parameter $\alpha$. In Section 3, we introduce the Fisher matrix analysis method and explain how we can transform from one set of variables to another set provided that we know how the variables depend on each other. We also describe the noise configuration for the BINGO telescope as well as some other important survey parameters. In Section 4, we present our results which can be divided into two sets. We first model the shift parameter $\alpha$ and some nuisance parameters assuming they can have different values at different redshift bins. We then use the Fisher matrix method to forecast constraints on the $\alpha$'s. The second set of results is our constraints on the cosmological parameters which we obtain by transforming our $\alpha$ Fisher matrix. We then discuss some issues with degeneracies and how we patch them with cosmic microwave background data. We make our conclusions in Section 5 and give some discussion on possible future work.

\section{BAO feature from the 21cm signal}
\label{sec:21cm_APS}

The 21cm brightness temperature perturbation in direction $\hat{n}$ is given in Hall et al.\cite{Hall_2013} 
\begin{equation}
\label{eq:brightness_temp_simp}
\begin{split}
\Delta_{T_b}(z,\hat{n}) & = \delta_n - \frac{1}{\mathcal{H}}\left[\hat{n} \cdot (\hat{n} \cdot \vec{\nabla})\vec{v}\right] \\ 
& + \left(\frac{d\ln(a^3\bar{n}_{\text{HI}})}{d\eta} - \frac{\dot{\mathcal{H}}}{\mathcal{H}} - 2\mathcal{H}\right)\delta\eta + \frac{1}{\mathcal{H}}\dot{\Phi} + \Psi,
\end{split}
\end{equation}
where $\delta_n$ is the HI density perturbation, $\mathcal{H}$ is the Hubble parameter in conformal time, $\vec{v}$ is the peculiar velocity, $\bar{n}_{\text{HI}}$ is the average HI number density, and $\Psi$ and $\Phi$ are the gravitational perturbation functions. Overdots indicate derivatives with respect to conformal time.

We then decompose this into spherical harmonics \cite{Hall_2013, BINGO_VII}
\begin{equation}
\label{eq:delta_T_b_sphere}
\Delta_{T_b}(z,\hat{n}) = \sum\limits_{lm} \Delta_{T_b,lm}(z)Y_{lm}(\hat{n}).
\end{equation}
We write the coefficients $\Delta_{T_b,lm}(z)$ in terms of the Fourier transforms of the perturbations
\begin{equation}
\label{eq:delta_T_b_sphere_coefficients}
\Delta_{T_b,lm}(z) = 4\pi i^l\int \frac{d^3\vec{k}}{(2\pi)^{3/2}}\Delta_{T_b,l}(\vec{k},z)Y_{lm}^*(\hat{k}),
\end{equation}
where
\begin{equation}
\label{eq:brightness_temp_fourier}
\begin{split}
\Delta_{T_b,l}(\Vec{k},z) & = \delta_nj_l(k\chi) + \frac{kv}{\mathcal{H}}j''_l(k\chi) +\biggl(\frac{1}{\mathcal{H}}\dot{\Phi} + \Psi\biggr)j_l(k\chi) \\
& - \biggl[\frac{1}{\mathcal{H}}\frac{d\ln(a^3\bar{n}_{\text{HI}})}{d\eta} - \frac{\dot{\mathcal{H}}}{\mathcal{H}^2} - 2\biggr] \\
& \times \biggl[\Psi j_l(k\chi) + vj'_l(k\chi) + \int_{0}^{\chi}(\dot{\Psi} + \dot{\Phi})j_l(k\chi')d\chi'\biggr],
\end{split}
\end{equation}
$j_l(k\chi)$ are the spherical Bessel functions, and primes on the Bessel functions denote derivatives with respect to the argument.

To get the data over some redshift bin we integrate over the redshift-normalized window function $W(z)$ \cite{BINGO_VII}
\begin{equation}
\label{delta_T_b_W}
\Delta_{T_b,l}^W(\vec{k}) = \int_{0}^{\infty} dz \bar{T}_b(z)W(z)\Delta_{T_b,l}(\vec{k},z).
\end{equation}
The angular power spectrum is then defined as \cite{Hall_2013}
\begin{equation}
\label{eq:C_l_21cm}
C_l^{WW'} = 4\pi \int d\ln k \mathcal{P}_{\mathcal{R}}(k)\Delta_{T_b,l}^W(\vec{k},z)\Delta_{T_b,l}^{W'}(\vec{k},z) \,,
\end{equation}
where $\mathcal{P}_{\mathcal{R}}$ is the dimensionless power spectrum of the primordial curvature perturbation $\mathcal{R}$.

\begin{figure}
\centering
\includegraphics[width=12cm]{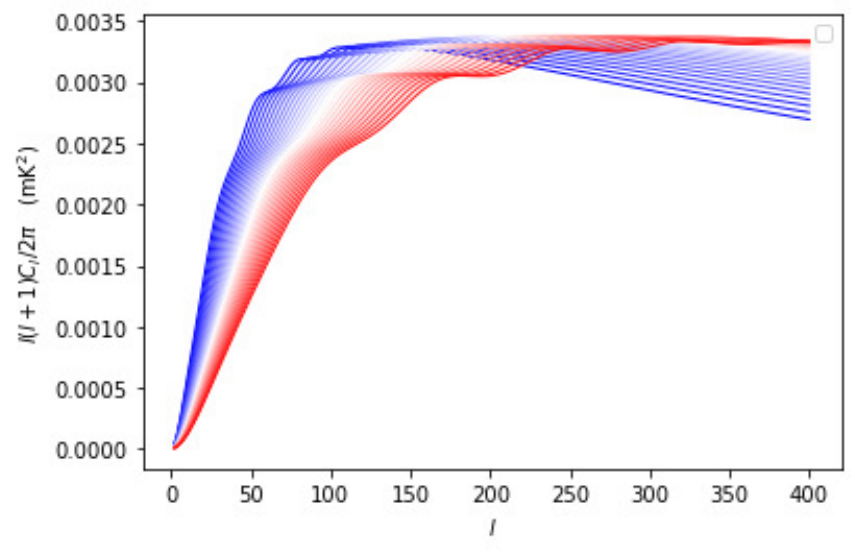}
\caption{Angular power spectra plotted over 30 redshift bins from $z = 0.13$ (leftmost blue line) to $z = 0.44$ (rightmost red line).}
\label{fig:C_ls}
\end{figure}

The biggest advantage of the 21cm $C_l$'s over CMB $C_l$'s is that the 21cm data is defined over a volume in redshift space, which allow for a tomographic analysis. More specifically we can explicitly see how the matter distribution of the universe evolves over time. Figure \ref{fig:C_ls} shows the 21cm $C_l$'s calculated for the 30 redshift bins used by the BINGO telescope. Going from higher redshifts (red) to lower redshifts (blue) we can observe three characteristics: first, as expected, matter structures grow and the power spectra increase especially at lower multipoles; second, the BAO scale moves to lower multipoles (larger angles), which means it increases for smaller redshifts; third, there is a suppression at higher multipoles. The BAO feature is easily identified as the ``wiggle" present across all redshifts.

While Eq.~(\ref{eq:C_l_21cm}) does give us the angular power spectrum for the 21cm distribution, we want a form of the APS that is given in terms of the matter power spectrum $P(k)$. Fortunately, this is easy to do with a few substitutions.

First, as noted in Hall et al.\cite{Hall_2013}, the first two terms in Eq.~(\ref{eq:brightness_temp_fourier}), related to the HI overdensity and redshift-space distortions, are several orders of magnitude larger than the other components. Therefore, we will restrict our calculations to these two contributions only. Second, we can solve the velocity in terms of the density, such that we rewrite the brightness temperature perturbation as \cite{Dodelson_2003}
\begin{equation}
\label{eq:delta_T_b_code}
\Delta_{T_b,l}(\vec{k},z) = \left( b_{\text{HI}}j_l(k\chi) - f(z)j''_l(k\chi)\right)D(z)\delta_m(0, k)  \,,
\end{equation}
where $D(z)$ and $f(z) \equiv d\ln{D(z)}/d\ln{a} $ are the growth function and growth rate, respectively. Taking the density component today outside $\Delta_{T_b,l}(\vec{k},z)$, we write the angular power spectra as \cite{BINGO_VIII}
\begin{equation}
\label{eq:C_l_code}
C_l^{WW'} = \frac{2}{\pi} \int dk \Delta_{T_b,l}^W(\vec{k},z)\Delta_{T_b,l}^{W'}(\vec{k},z)k^2P(k) \,,
\end{equation}
where $P(k)$ is the $3D$ matter power spectrum today, which is related to the transfer function $T(k)$ by \cite{Bunn_1997} 
\begin{equation}
\label{eq:Delta^2(k)}
P(k) = \delta_H^2\frac{2\pi^2}{k^3}\left(\frac{ck}{H_0}\right)^{3+n}T^2(k),
\end{equation}
where $\delta_H$ is the density perturbation at horizon crossing.

\begin{figure}[t]
\centering
\includegraphics[width=11cm]{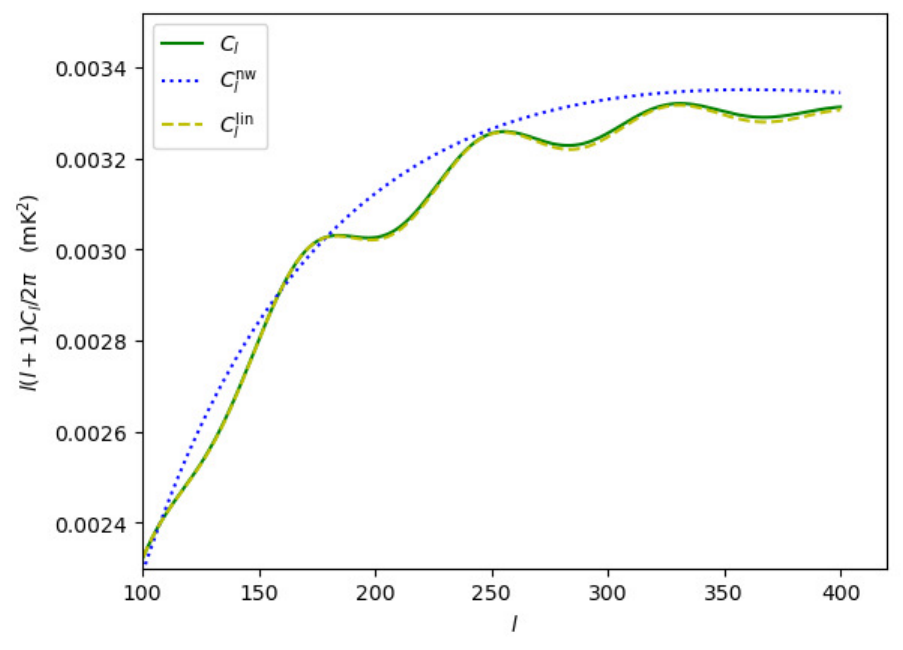}
\includegraphics[width=11cm]{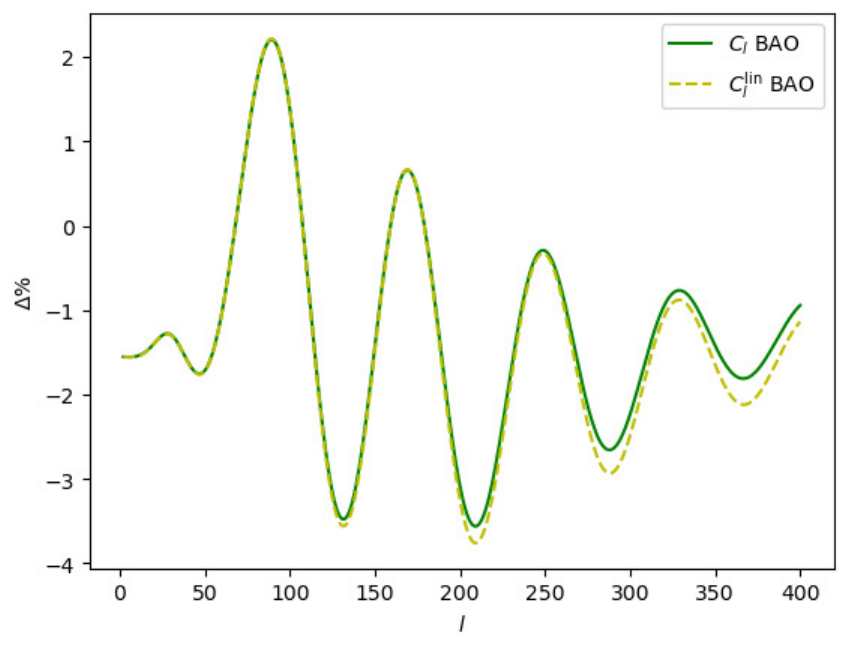}
\caption{(Top panel) Comparison of $C_l^{\text{temp}}$ (green solid), $C_l^{\text{nw}}$ (blue dotted), and $C_l^{\text{lin}}$ (yellow dashed). The difference between $C_l^{\text{temp}}$ and $C_l^{\text{lin}}$ is the nonlinear damping which suppresses the oscillations at higher multipoles. (Bottom panel) BAO scale showing the percentage deviation of the full and linear APS from the no-wiggle APS. }
\label{fig:Cls_comp}
\end{figure}

We use a numerical fit for the transfer function as given by Eisenstein and Hu\cite{Eisenstein_1998}. This fit works in the linear regime, thus the power spectra is linear as well. To account for the growth of nonlinear structure, we make a template power spectrum \cite{BINGO_VIII, Camacho_2019}
\begin{equation}
\label{eq:P_temp(k)}
P_{\text{temp}}(k) = [P_{\text{lin}}(k) - P_{\text{nw}}(k)]e^{-k^2\Sigma_{nl}^2} + P_{\text{nw}}(k),
\end{equation}
where $\Sigma_{nl} = 5.2$Mpc/h is the nonlinear damping scale given by Chan, et al.\cite{Chan_2018}. $P_{\text{lin}}(k)$ refers to the full linear fit and $P_{\text{nw}}(k)$ is the fit where the BAOs have been removed. We use this equation for $P(k)$ in Eq.~(\ref{eq:C_l_code}). Because the Fisher matrix method only forecasts standard deviations and not expectation values, we must obtain our fiducial parameters from an outside source. We use the values obtained by the $Planck$ satellite shown in the first column of Table 1 in $Planck$\cite{2020}. For convenience we rewrite these values in Table \ref{tab:params}. Note that we also insert the fiducial values for the CPL parameterization as well as our template parameters $B_0$, $A_i$, and $\alpha$. Figure \ref{fig:Cls_comp} shows the $C_l$'s obtained from inserting $P_{\text{temp}}(k)$, $P_{\text{lin}}$, and $P_{\text{nw}}$ into Eq. (\ref{eq:C_l_code}), and also the effect of the template on the BAO scale alone.

We intend to extract from the 21cm intensity maps the scale associated with the BAO feature, which is encoded in the angular diameter distance $d_A(z)$. Following Camacho et al.\cite{Camacho_2019}, we use a template method and write the power spectrum signal as
\begin{equation}
\label{eq:C_l_temp}
C(l) = B_0C_{\text{temp}}(l/\alpha) + A_0 + A_1l + A_2/l^2,
\end{equation}
where $B_0$ and $A_i$ are nuisance parameters. Specifically, the amplitude $B_0$ is related to the linear bias squared and the parameters $A_i$ account for scale-dependent bias, shot noise, uncertainties in the redshift space distortion, etc. We set the fiducial values as $B_0 = 1$ and $A_i = 0$, and allow them to vary with redshift.

Most importantly, Eq. (\ref{eq:C_l_temp}) has a much more powerful feature than just allowing us to account for more sources of error. The shift parameter $\alpha$ measures the shift of the BAO peak positions with respect to our fiducial cosmology. It is defined as \cite{Camacho_2019}
\begin{equation}
\label{eq:alpha}
\alpha \equiv \frac{(d_A(z)/r_s)}{(d_A(z)/r_s)}_{\text{fid}},
\end{equation}
where $d_A(z)$ is the angular diameter distance and $r_s$ is the sound horizon at drag epoch. Measuring $\alpha$ is a very good test for how accurate a fiducial cosmology is. Not only that, but $\alpha$ only depends on the BAO effect, it is completely independent of bias and other parameters meaning that we only have to understand BAO physics well to get accurate cosmological constraints.

\section{Fisher Matrix Analysis}\label{sec:fisher}

For a set of parameters $\theta_i$ and a model $\mathcal{M}$, the Fisher information matrix can be calculated as \cite{Tegmark2_1997}
\begin{equation}
\label{eq:F}
F_{ij} \equiv \biggl\langle \frac{\partial^2\mathcal{L}}{\partial\theta_i\partial\theta_j} \biggl\rangle = \frac{1}{2}\mathrm{Tr}\left[\mathrm{\bf C}^{-1}\frac{\partial\mathrm{\bf C}}{\partial\theta_i}\mathrm{\bf C}^{-1}\frac{\partial\mathrm{\bf C}}{\partial\theta_j}\right] \,,
\end{equation}
where $\mathcal{L} = -\ln{L}$ and $L$ is the likelihood function. 

We write the covariance $\textbf{C}$ as the sum of the template $C_l$s and the noise spectra  
\begin{equation}
\label{C_l_tot}
\textbf{C} = C(l) + C_l^{\text{shot}}(z_i,z_j) + N_l(z_i,z_j) \,,
\end{equation}
where $C_l^{\text{shot}}(z_i,z_j)$ is the shot noise spectrum given by \cite{Hall_2013,BINGO_VII}
\begin{equation}
C_l^{\text{shot}}(z) = \frac{\bar{T}^2(z)}{\bar{N}(z)} \,,
\end{equation}
with $\bar{N}(z)$ the angular density of the sources and $\bar{T}$ the average of 21cm brightness temperature. The thermal noise $N_l$ can be calculated as given in Costa et al\cite{BINGO_VII}
\begin{equation}
N_l = \frac{T_{\text{sys}}^2}{\Delta\nu t_{\text{obs}}}\left(\frac{4\pi f_{\text{sky}}}{2n_f}\right),
\end{equation}
where $f_{\text{sky}}$ is the surveyed fraction of the sky and it is assumed that two polarizations are measured. Since BINGO is a fixed dish telescope, it surveys the sky that passes above it through Earth's rotation. Additionally, parts of the survey range that intersect with the galactic plane are masked as the noise is too bright in those regions. Our noise simulation uses a survey fraction of $f_{\text{sky}} = 7\%$ \cite{BINGO_V}. We also take into account the maximum beam resolution of the telescope by multiplying the thermal noise by an exponential factor. We use the same noise values as in BINGO VII, which also investigates the effects of noise on the signal. Table \ref{tab:BINGO_telescope} gives the fiducial parameters for the BINGO survey.

\begin{table}[h]
    \centering
    \tbl{Fiducial parameters of the BINGO telescope \cite{BINGO_VII}.}
    {\begin{tabular}{l c}
        \hline
        Parameter & BINGO value \\
        \hline
        Frequency range & $[980, 1260]$ MHz \\
        Redshift range & $[0.127, 0.0449]$ \\
        Number of frequency channels & 30 \\
        Number of feed horns & 28 \\
        Sky coverage with galactic mask & 2900 $\text{deg}^2$ \\
        Observational time ($t_{\text{obs}}$) & 1 year \\
        System temperature ($T_{\text{sys}}$) & 70 K \\
        Beam resolution ($\theta_{\text{FWHM}}$) & 40 arcmin \\
        \hline
    \end{tabular}\label{tab:BINGO_telescope}}

\end{table}

We follow a similar approach to the CMB case, where we decompose the brightness temperature perturbation in spherical harmonics \cite{BINGO_VIII}. We note however that the CMB signal comes from one specific redshift and yields a 2-dimensional map, while 21cm surveys spans a whole volume and are 3-dimensional with multiple redshifts. Then, the CMB diagonal matrix is transformed into the block diagonal matrix \cite{BINGO_VII}
\begin{equation}
\label{eq:C_block}
C =
\begin{bmatrix}
A_{l=2} & 0 & \dots & 0 \\
0 & A_3 & \dots & 0 \\
\vdots & \vdots & \dots & \vdots \\
0 & 0 & \dots & A_n 
\end{bmatrix},
\end{equation}
where 
\begin{equation}
\label{eq:A_block}
A_l = (2l + 1)
\begin{bmatrix}
C_l(z_1,z_1) & C_l(z_1,z_2) & \dots & C_l(z_1,z_n) \\
C_l(z_2,z_1) & C_l(z_2,z_2) & \dots & C_l(z_2,z_n) \\
\vdots & \vdots & \dots & \vdots \\
C_l(z_n,z_1) & C_l(z_n,z_2) & \dots & C_l(z_n,z_n) 
\end{bmatrix}.
\end{equation}

Finally, if we have a Fisher matrix, $F_{ij}$, defined in terms of variables $\vec{\Theta} = (\theta_1, \theta_2,\dots, \theta_n)$, we can transform to a new Fisher matrix, $F'_{kl}$, defined in terms of variables $\vec{\Theta}' = (\theta'_1, \theta'_2,\dots, \theta'_m)$ for $m \le n$ using the equation \cite{coe2009fisher}
\begin{equation}
\label{eq:F_kl}
F'_{kl} = \sum_{ij} \frac{\partial \theta_i}{\partial \theta'_k}\frac{\partial \theta_j}{\partial \theta'_l}F_{ij}.
\end{equation}

\section{Results}

\begin{table}
\centering
\tbl{ Fiducial values for cosmological parameters. $b_{\text{HI}}$ and $\Omega_{\text{HI}}$ are related to 21cm physics. $B_0$, $A_i$, and $\alpha$ are the $C_l$ template parameters.}
{\begin{tabular}{l c}
\hline
Parameter & Fiducial Value \\ \hline
$\Omega_bh^2 $            & 0.022383          \\
$\Omega_ch^2$             & 0.12011           \\
$h$                       & 0.6732            \\
$n_s$                     & 0.96605           \\
$A_s$                     & $2.1\times 10^{-9}$ \\
$w_0$                     &  -1               \\
$w_a$                     & 0 \\
$b_{\text{HI}}$           &   1  \\
$\Omega_{\text{HI}}$      & $6.2 \times 10^{-4}$ \\
$B_0$                     & 1 \\
$A_i$                     & 0 \\
$\alpha$                  & 1 \\
\hline
\end{tabular}\label{tab:params}}
\end{table}

We make our analysis in two steps. First, we calculate the Fisher matrix using the covariance for the template angular power spectra given by Eqs.~(\ref{C_l_tot}) and (\ref{eq:C_l_temp}). We take the BAO shift parameter $\alpha$ and the nuisance parameters $B_0$ and $A_i$, and model a redshift dependence by assuming an independent constant every three redshift bins over the 30 bins of our survey. This gives us a total of 50 parameters, 10 useful $\alpha$'s and 40 nuisance parameters. We do this for two reasons. Firstly, allowing the $\alpha$'s to assume different values in different bins makes it possible to study its evolution with redshift. The second reason is that we ultimately want to transform our Fisher matrix of template variables into a Fisher matrix of physical variables given in Table \ref{tab:params}. Since we can only transform Fisher matrices into other Fisher matrices of the same size or smaller, we must ensure we start out with enough parameters. Splitting the parameters across redshift bins not only gives us more parameters. It also allows us to investigate how these parameters vary over time. We choose to share each $\alpha$ and nuisance parameter over three redshift bins to save computation time as the time required to calculate an $n \times n$ Fisher matrix goes like $~ \frac{1}{2} n^2$. It also allows an easy comparison to BINGO VIII which uses the same sharing scheme.

We use these 50 parameters to calculate a $50\times50$ Fisher matrix, which can be done using Eq.~(\ref{eq:F}), setting all $\alpha$'s, $B_0$'s, and $A_i$'s to their fiducial values, and calculating the derivatives numerically. This is the most computationally intensive step in our analysis by far. Since $B_0$ and the $A_i$'s are nuisance parameters, we marginalize over them by deleting their corresponding rows and columns in the covariance matrix (the covariance matrix is simply the matrix inverse of the Fisher matrix). Taking the square root of the diagonal entries in the covariance matrix gives us the standard deviation of all 10 $\alpha$'s. In order to compare our result with BINGO VIII we average these 10 $\sigma_{\alpha}$'s across three redshift bins to get the results shown in Table \ref{tab:alphas}. For comparison, we also include $\sigma_{\alpha}$'s from the Markov Chain Monte Carlo method from BINGO VIII.

\begin{table}
\centering
\tbl{Standard deviations for $\alpha$ across three redshift bins. Markov Chain Monte Carlo results are from BINGO VIII.}
{\begin{tabular}{l c c}
\hline
z-bins & $\langle \sigma_{\alpha} \rangle$ Fisher  & $\langle \sigma_{\alpha} \rangle$ MCMC\\ 
\hline
1-10   & 0.0816 & 0.0653              \\ 
11-20  & 0.0636 & 0.0410               \\ 
21-30  & 0.0508 & 0.0322               \\ 
\hline
\end{tabular}\label{tab:alphas}}
\end{table}

Our analysis so far has been similar to what was done in BINGO VIII. We go further by transforming our forecasts in the BAO shift parameters into forecasts in the final cosmological parameters. As we saw in Eq.~(\ref{eq:alpha}), the shift parameter $\alpha$ depends on the angular diameter distance and the sound horizon at the drag epoch, which are ultimately functions of $\Omega_bh^2$, $\Omega_ch^2$, $h$, and the dark energy equation of state parameters $w_0$ and $w_a$. Therefore, the uncertainties in $\alpha$s can be related to uncertainties in these parameters, which can be done by transforming our $10 \times 10 \; \alpha$ Fisher matrix into a Fisher matrix for our final cosmological parameters.

 It can be shown that the BAOs are insensitive to the bias $b_{\text{HI}}$ and the neutral hydrogen density $\Omega_{\text{HI}}$ and therefore, the $\alpha$'s are also insensitive as well. Thus, errors in these parameters have no effect on the physical parameters. $\alpha$ is also insensitive to $A_s$ and $n_s$ further reducing the potential for systemic errors to affect the analysis.

\begin{table}[t]
\centering
\tbl{Cosmological Forecasts for $\Lambda$CDM.}
{\begin{tabular}{l c c c}
\hline
& $Planck$ & BINGO VII + $Planck$ & BINGO BAO + $Planck$ \\ \hline
Parameter & $\pm1\sigma\;(100\%\times\sigma/ \theta^{fid}_i)$      & $\pm1\sigma\;(100\%\times\sigma/ \theta^{fid}_i)$                & $\pm1\sigma\;(100\%\times\sigma/ \theta^{fid}_i)$              \\ \hline
$\Omega_bh^2$     & 0.00015 (0.7\%) & 0.00013 (0.6\%) & 0.00014 (0.6\%) \\ 
$\Omega_ch^2$     & 0.0014  (1.1\%) & 0.0010  (0.8\%) & 0.0012 (1.0\%) \\ 
$h$               & 0.0061  (0.9\%) & 0.0045  (0.7\%) & 0.0055 (0.8\%) \\ 
$\ln(10^{10}A_s)$ & 0.016   (0.5\%) & 0.015   (0.5\%) & 0.016 (0.5\%) \\ 
$n_s$             & 0.0043  (0.4\%) & 0.0039  (0.4\%) & 0.0041 (0.4\%) \\
\hline
\end{tabular}\label{tab:constraints_lcdm}}
\end{table}

\begin{table}[h]
\centering
\tbl{Cosmological Forecasts for $w$CDM.}
{\begin{tabular}{l c c c}
\hline
& $Planck$ & BINGO VII + $Planck$ & BINGO BAO + $Planck$ \\ \hline
Parameter & $\pm1\sigma\;(100\%\times\sigma/ \theta^{fid}_i)$      & $\pm1\sigma\;(100\%\times\sigma/ \theta^{fid}_i)$                & $\pm1\sigma\;(100\%\times\sigma/ \theta^{fid}_i)$              \\ \hline
$\Omega_bh^2$     & 0.00015 (0.7\%) & 0.00014 (0.6\%) & 0.00015(0.7\%) \\ 
$\Omega_ch^2$     & 0.0014  (1.2\%) & 0.0011  (0.9\%) & 0.0014 (1.2\%) \\ 
$h$               & 0.089    (13\%) & 0.0073  (1.1\%) & 0.020 (2.9\%)  \\ 
$\ln(10^{10}A_s)$ & 0.016   (0.5\%) & 0.016   (0.5\%) & 0.0016 (0.5\%) \\ 
$n_s$             & 0.0044  (0.5\%) & 0.0040  (0.4\%) & 0.0043 (0.4\%) \\
$w_0$             & 0.26     (25\%) & 0.033   (3.3\%) & 0.075 (7.5\%)  \\
\hline
\end{tabular}\label{tab:constraints_wcdm}}
\end{table}

\begin{table}[h]
\centering
\tbl{Cosmological Forecasts for CPL.}
{\begin{tabular}{l c c c}
\hline
& $Planck$ & BINGO VII + $Planck$ & BINGO BAO + $Planck$ \\ \hline
Parameter & $\pm1\sigma\;(100\%\times\sigma/ \theta^{fid}_i)$      & $\pm1\sigma\;(100\%\times\sigma/ \theta^{fid}_i)$                & $\pm1\sigma\;(100\%\times\sigma/ \theta^{fid}_i)$              \\ \hline
$\Omega_bh^2$     & 0.00016 (0.7\%) & 0.00014 (0.6\%) & 0.00015 (0.7\%) \\ 
$\Omega_ch^2$     & 0.0013  (1.1\%) & 0.0011  (0.9\%) & 0.0013 (1.1\%)  \\ 
$h$               & 0.088    (13\%) & 0.019   (2.9\%) & 0.029 (4.4\%)   \\ 
$\ln(10^{10}A_s)$ & 0.016   (0.5\%) & 0.016   (0.5\%) & 0.016 (0.5\%)   \\ 
$n_s$             & 0.0044  (0.5\%) & 0.0041  (0.4\%) & 0.0044 (0.5\%)  \\
$w_0$             & 0.46     (46\%) & 0.30     (30\%) & 0.40 (40\%)     \\
$w_a$             & 1.8             & 1.2             & 1.7             \\
\hline
\end{tabular}}\label{tab:constraints_cpl}
\end{table}

Unfortunately, as discussed in Adame et al. \cite{desicollaboration2024desi}, there is a large degeneracy between $H_0$ and $r_d$ using the BAO data only, and we can only determine the combination $H_0r_s$. In order to break this degeneracy, we need prior knowledge about $\Omega_bh^2$. This can be done by combining the BAO data with some external data which properly constrain this parameter. Therefore, we combine our results from the 21cm intensity mapping with the obtained from CMB data by the $Planck$ satellite. We do this for three cosmic acceleration models $\Lambda$CDM, $w$CDM, and CPL ($w_0w_a$CDM). Our results are shown in Tables~4,5, and 6 for $\Lambda$CDM, $w$CDM, and CPL, respectively. We note that the parameters $A_s$ and $n_s$ are included in the tables to account for their degeneracies within the $Planck$ data, and the improvements on these constraints are minuscule.

\begin{figure}[t]
\centering
\includegraphics[width=12cm]{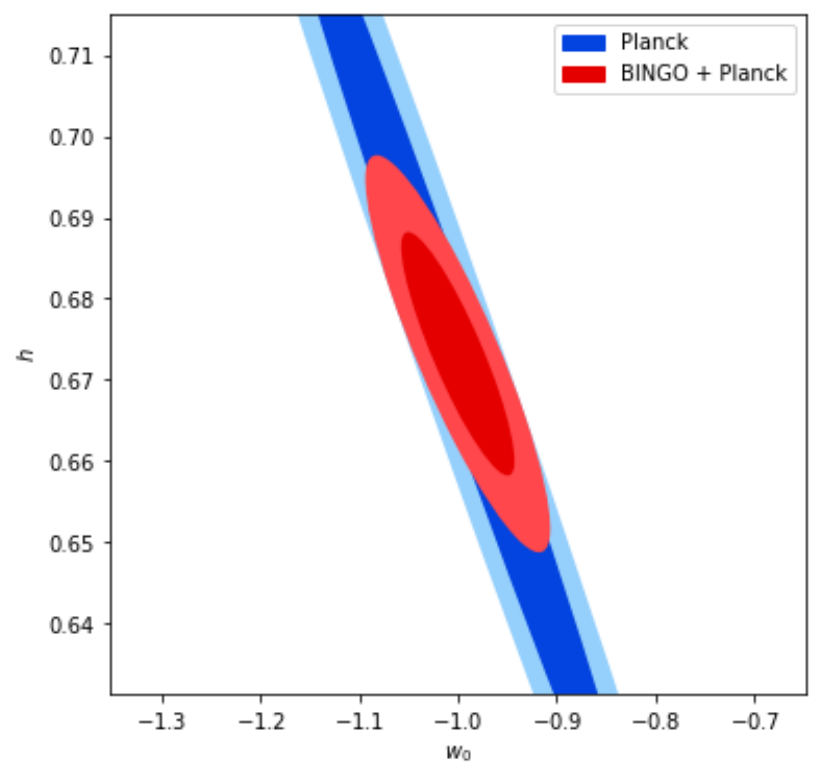}
\caption{Marginalized $68\%$ and $95\%$ Confidence Ellipses for $w_0$ and $h$ in the $w$CDM model.}
\label{fig:ellipse}
\end{figure}

\begin{figure}[t]
\centering
\includegraphics[width=12cm]{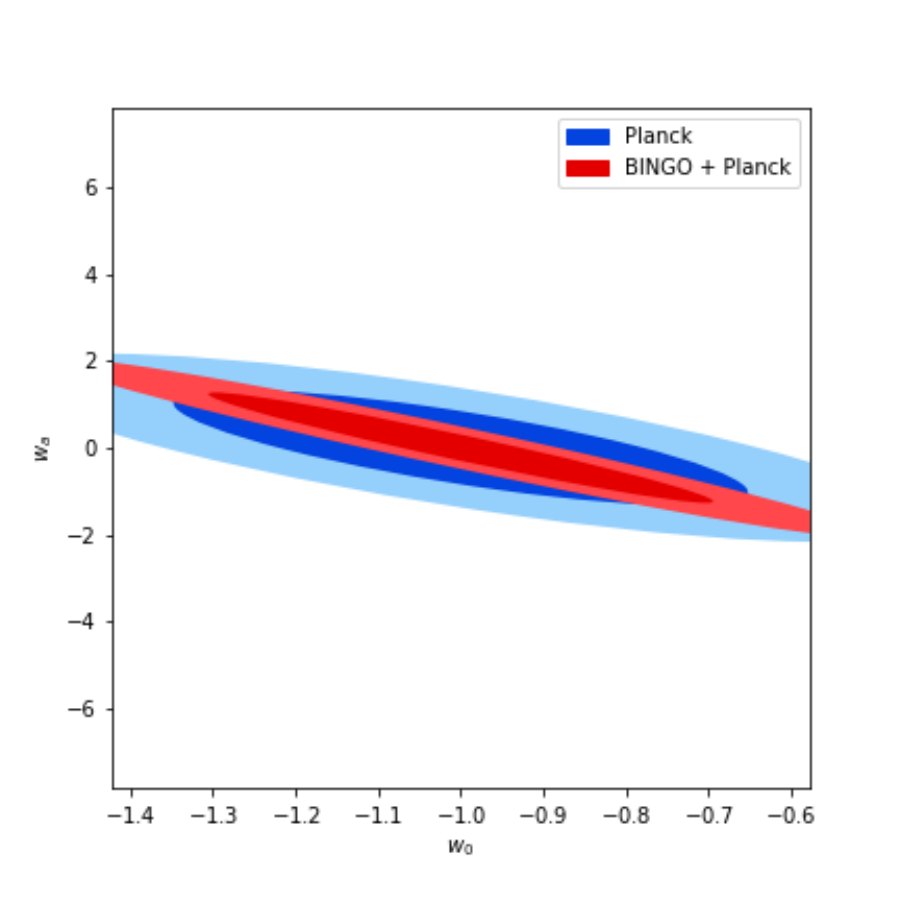}
\caption{Marginalized $68\%$ and $95\%$ Confidence Ellipses for $w_0$ and $w_a$ in the CPL model.}
\label{fig:ellipse2}
\end{figure}

In most cases, the combined BINGO BAO + $Planck$ constraints are not statistically any stronger than the constraints from $Planck$ alone. However, the constraints for $h$, $w_0$, and $w_a$ do get improved significantly over $Planck$. This makes sense as more data related to cosmic expansion is encoded in late-time 21cm data than in early-time CMB data. Unfortunately, these constraints are not as strong as those obtained by calculating the Fisher matrix from the entire 21cm angular power spectrum as done in BINGO VII \cite{BINGO_VII}. However, as discussed above, we expect our analysis to be less susceptible to systematic effects related with uncertainties on HI variables. Figure \ref{fig:ellipse} shows the confidence ellipses for $w_0$ and $h$ in the $w$CDM model for both $Planck$ and BINGO BAO + $Planck$. Figure \ref{fig:ellipse2} shows the confidence ellipses for $w_0$ and $w_a$ in the CPL model.

\section{Discussions and Conclusions}\label{sec:conclusions}

We have found that calculating a Fisher matrix of a template power spectrum is capable of forecasting constraints on the BAO peak shift parameter $\alpha$ with results comparable to, but slightly less optimistic than, the MCMC analysis in BINGO VIII \cite{BINGO_VIII}. It is not clear for us why the MCMC result produces better results in this case, but this makes our analysis more conservative. Unlike the MCMC analysis, the Fisher matrix method does not calculate the best fit values for the parameters, however, it is significantly less computationally intensive.

We have also found that transforming this Fisher matrix into a Fisher matrix of cosmological parameters is far too degenerate to give any useful forecasts on its own. It has actually been demonstrated, e.g. in the DESI collaboration \cite{desicollaboration2024desi}, that the only cosmological forecasts that can be obtained from BAO alone in a flat $\Lambda$CDM model are in the $\Omega_m$ and $H_0r_d$. Because of this, we combine the BAO information with CMB data to introduce a prior in $\Omega_bh^2$ and break those degeneracies. Our results show that the BAO data will mainly constrain the Hubble parameter $h$ and the dark energy equation of state paramters $w_0$ and $w_a$. In fact, the best results were obtained for the $w$CDM model, which improved the constraints of the $Planck$ satellite from $h = 0.089 \, (13\%)$ and $w_0 = 0.26 \, (25\%)$ to $h = 0.020 \, (2.9\%)$ and $w_0 = 0.075 \, (7.5\%)$. This result is consistent with what was obtained by Olivari et al. \cite{Olivari:2017bfv}, although they used a different methodology to extract the BAO information. 

We also compare our analysis with the results from the full angular power spectra obtained in BINGO VII \cite{BINGO_VII}. As expected, the full power spectra contains more information than the BAO wiggles alone. Therefore, our constraints were at the same level or worse than the previous analysis. However, the BAO do not depend on the overall amplitude of the 21cm angular power spectra, which makes it less susceptible to systematic effects from foregrounds removal, bias, and non-linearities. We note that both the full APS BINGO + $Planck$ and the BINGO BAO + $Planck$ data provide weaker forecasts than the recently published results from the DESI project\cite{desi2025}. There are a few reasons why this could be. It could have something to do with the method we are using to extract the BAO from the APS, but that wouldn't explain why DESI + $Planck$ is better than BINGO APS + $Planck$. We also note that because both the BINGO APS and the BINGO BAO use different methodology from DESI, there are different systematic effects that could be investigated more rigorously in future works. Finally, we must remember that because BINGO is still under construction, all the work we have done so far is produce forecasts while DESI has already released it observational results. BAO have never been detected using 21cm intensity mapping so there are likely some factors that haven't been anticipated. It is the authors opinion that 21cm APS data not only from BINGO, but from other projects such as SKA will be complementary to spectroscopic data from projects like DESI. The relatively simple design of the BINGO instrument makes it ideal for testing new ideas that could be applied to future surveys.

There is some potential for future work using similar methodology. In particular, we used the Fisher matrix analysis method to obtain our $\alpha$ forecasts. While this method is quick, it only forecasts standard deviations and not expected values. This means that we lose some accuracy as we must make assumptions on the fiducial values for variables. This is generally not that much of a disadvantage for cosmological variables as we have used measured values from $Planck$ as our fiducial values. However, we used the fiducial value of $1$ for all of our $\alpha$'s. While this is what we would expect given the definition of $\alpha$, we would obtain better feedback on our model if we allowed the $\alpha$'s to vary. Using the results of BINGO VIII we could input the MCMC $\alpha$ expectation values as our fiducial values for the $\alpha$'s in our Fisher analysis and see how this affects our results.

It must be noted that in this analysis, we only used the angular BAO signal and ignored the radial BAO signal. We chose to do this because we used the angular power spectrum as used in BINGO VII and BINGO VIII which does not include radial information. As we did calculate the APS for each of our 30 redshift bins, we do have radial data, but our analysis method only considers the angular BAO signal as the $\alpha$'s for each redshift bin are calculated independently of each other. Many past works such as Villaescusa-Navarro, Alonso \& Viel 2017 \cite{Villaescusa_Navarro_2016}, Kennedy \& Bull 2021 \cite{10.1093/mnras/stab1814}, and Avila et al. 2022 \cite{avila_2022} have focused on the radial BAO signal in the context of 21cm intensity mapping. Of particular note, these studies have found that the radial BAO signal is actually more robust for 21cm intensity mapping. Since we allow $\alpha$ to vary across every three redshift bins, there is potential for extracting radial BAO information, especially if ones allows $\alpha$ to vary across more redshift bins in the BINGO data. Other radio telescopes with more redshift bins and/or a larger redshift range would give more radial information and likely lead to more precise data on large scale structure formation and late-time cosmic acceleration.

\section*{Acknowledgments}

Supported by the National Natural Science Foundation of China (12175192, 12005184).Andr\'{e} A. Costa also acknowledges financial support from the Conselho Nacional de Desenvolvimento Cient\'{i}fico e Tecnol\'{o}gico (CNPq) - Brazil (grant 102734/2024-0).


\begin{thebibliography}{0}
\bibitem{pe00300h}
P.~J.~E. Peebles and J.~T. Yu.
\newblock Primeval adiabatic perturbation in an expanding universe.
\newblock {\em Astrophysical Journal}, 162:815--836, 1970.

\bibitem{Eisenstein_2007}
Daniel~J. Eisenstein, Hee-Jong Seo, and Martin White.
\newblock On the robustness of the acoustic scale in the low-redshift
  clustering of matter.
\newblock {\em The Astrophysical Journal}, 664(2):660--674, August 2007.

\bibitem{1968ApJ...151..459S}
Joseph Silk.
\newblock {Cosmic black body radiation and galaxy formation}.
\newblock {\em Astrophys. J.}, 151:459--471, 1968.

\bibitem{PhysRevLett.5.361}
H.~M. Goldenberg, D.~Kleppner, and N.~F. Ramsey.
\newblock Atomic hydrogen maser.
\newblock {\em Phys. Rev. Lett.}, 5:361--362, 10 1960.

\bibitem{2000AJ....120.1579Y}
Donald~G. York et~al.
\newblock {The Sloan Digital Sky Survey: Technical Summary}.
\newblock {\em Astron. J.}, 120:1579--1587, 2000.

\bibitem{Amiri_2018}
M.~Amiri et~al.
\newblock {The CHIME Fast Radio Burst Project: System Overview}.
\newblock 3 2018.

\bibitem{MeerKLASS:2017vgf}
Mario~G. Santos et~al.
\newblock {MeerKLASS: MeerKAT Large Area Synoptic Survey}.
\newblock In {\em {MeerKAT Science}: {On the Pathway to the SKA}}, 9 2017.

\bibitem{2020PASA...37....2W}
A.~Weltman et~al.
\newblock {Fundamental physics with the Square Kilometre Array}.
\newblock {\em Publ. Astron. Soc. Austral.}, 37:e002, 2020.

\bibitem{Battye_2013}
R.~A. Battye, et~al.
\newblock Hi intensity mapping: a single dish approach.
\newblock {\em Monthly Notices of the Royal Astronomical Society},
  434(2):1239--1256, July 2013.

\bibitem{10.1093/mnras/stv2153}
M.-A. Bigot-Sazy, et~al.
\newblock {Simulations for single-dish intensity mapping experiments}.
\newblock {\em Monthly Notices of the Royal Astronomical Society},
  454(3):3240--3253, 10 2015.

\bibitem{BINGO_I}
Elcio Abdalla et~al.
\newblock {The BINGO project - I. Baryon acoustic oscillations from integrated
  neutral gas observations}.
\newblock {\em Astron. Astrophys.}, 664:A14, 2022.

\bibitem{BINGO_II}
Carlos~A. Wuensche et~al.
\newblock {The BINGO project - II. Instrument description}.
\newblock {\em Astron. Astrophys.}, 664:A15, 2022.

\bibitem{BINGO_III}
Filipe~B. Abdalla et~al.
\newblock {The BINGO Project - III. Optical design and optimization of the
  focal plane}.
\newblock {\em Astron. Astrophys.}, 664:A16, 2022.

\bibitem{BINGO_IV}
Vincenzo Liccardo et~al.
\newblock {The BINGO project - IV. Simulations for mission performance
  assessment and preliminary component separation steps}.
\newblock {\em Astron. Astrophys.}, 664:A17, 2022.

\bibitem{BINGO_V}
Karin S.~F. Fornazier et~al.
\newblock {The BINGO project - V. Further steps in component separation and
  bispectrum analysis}.
\newblock {\em Astron. Astrophys.}, 664:A18, 2022.

\bibitem{BINGO_VI}
Jiajun Zhang et~al.
\newblock {The BINGO project - VI. H I halo occupation distribution and mock
  building}.
\newblock {\em Astron. Astrophys.}, 664:A19, 2022.

\bibitem{BINGO_VII}
Andre~A. Costa et~al.
\newblock {The BINGO project - VII. Cosmological forecasts from 21 cm intensity
  mapping}.
\newblock {\em Astron. Astrophys.}, 664:A20, 2022.

\bibitem{BINGO_VIII}
Camila~Paiva Novaes et~al.
\newblock {The BINGO project - VIII. Recovering the BAO signal in HI intensity
  mapping simulations}.
\newblock {\em Astron. Astrophys.}, 666:A83, 2022.

\bibitem{Hall_2013}
Alex Hall, Camille Bonvin, and Anthony Challinor.
\newblock Testing general relativity with 21-cm intensity mapping.
\newblock {\em Physical Review D}, 87(6), March 2013.

\bibitem{Dodelson_2003}
Scott Dodelson.
\newblock {\em Modern cosmology}.
\newblock Academic Press, 2003.

\bibitem{Bunn_1997}
Emory~F. Bunn and Martin White.
\newblock The 4 yearcobenormalization and large-scale structure.
\newblock {\em The Astrophysical Journal}, 480(1):6--21, May 1997.

\bibitem{Eisenstein_1998}
Daniel~J. Eisenstein and Wayne Hu.
\newblock Baryonic features in the matter transfer function.
\newblock {\em The Astrophysical Journal}, 496(2):605--614, April 1998.

\bibitem{Camacho_2019}
H.~Camacho et~al.
\newblock {Dark Energy Survey Year 1 Results: Measurement of the Galaxy Angular
  Power Spectrum}.
\newblock {\em Mon. Not. Roy. Astron. Soc.}, 487(3):3870--3883, 2019.

\bibitem{Chan_2018}
K.~C. Chan et~al.
\newblock {BAO from Angular Clustering: Optimization and Mitigation of
  Theoretical Systematics}.
\newblock {\em Mon. Not. Roy. Astron. Soc.}, 480(3):3031--3051, 2018.

\bibitem{Tegmark2_1997}
Max Tegmark, Andy~N. Taylor, and Alan~F. Heavens.
\newblock Karhunen-loeve eigenvalue problems in cosmology: How should we tackle
  large data sets?
\newblock {\em The Astrophysical Journal}, 480(1):22--35, May 1997.

\bibitem{coe2009fisher}
Dan Coe.
\newblock Fisher matrices and confidence ellipses: A quick-start guide and
  software, 2009.

\bibitem{2020}
N.~Aghanim et~al.
\newblock {Planck 2018 results. VI. Cosmological parameters}.
\newblock {\em Astron. Astrophys.}, 641:A6, 2020.
\newblock [Erratum: Astron.Astrophys. 652, C4 (2021)].

\bibitem{desicollaboration2024desi}
A.~G. Adame et~al.
\newblock {DESI 2024 VI: Cosmological Constraints from the Measurements of Baryon Acoustic Oscillations}.
\newblock 4 2024.

\bibitem{Olivari:2017bfv}
L.~C. Olivari, et al.
\newblock {Cosmological parameter forecasts for HI intensity mapping
  experiments using the angular power spectrum}.
\newblock {\em Mon. Not. Roy. Astron. Soc.}, 473(3):4242--4256, 2018.

\bibitem{Villaescusa_Navarro_2016}
Villaescusa-Navarro, et~al.
\newblock{Baryonic acoustic oscillations from 21cm intensity mapping: the Square Kilometre Array case}
\newblock{\em Mon. Not. Roy. Astron. Soc.}, 466(3):2736–2751, 2016.

\bibitem{10.1093/mnras/stab1814}
Kennedy, et~al.
\newblock{Statistical recovery of the BAO scale from multipoles of the beam-convolved 21cm correlation function}
\newblock{\em Mon. Not. Roy. Astron. Soc.}, 506(2):2638-2658, 2021

\bibitem{avila_2022}
Avila, et~al.
\newblock HI IM correlation function from UNIT simulations: BAO and observationally induced anisotropy
\newblock \em{Mon. Not. Roy. Astron. Soc.}, 510(1):292-308. 2022

\bibitem{desi2025}
Abdul-Karim, et al.
\newblock{
DESI DR2 Results II: Measurements of Baryon Acoustic Oscillations and Cosmological Constraints}
\newblock, 2025

\end{thebibliography}
\end{document}